\documentclass[12pt]{article}
\usepackage{graphicx}
\usepackage{epstopdf}

\newcommand{\pt}{$p_\mathrm{T}$}
\newcommand{\kt}{$k_\mathrm{T}$}
\newcommand{\bb}{$b\bar{b}$ }

\newcommand{\pp}{$p\bar{p}$ }

\def\Title#1{\begin{center} {\Large #1 } \end{center}}
\def\Author#1{\begin{center}{ \sc #1} \end{center}}
\def\Address#1{\begin{center}{ \it #1} \end{center}}

\newenvironment{Abstract}{\begin{center}{\bf Abstract}\end{center} \bigskip \begin{quotation}}{\end{quotation}}
\newenvironment{Presented}{\begin{quotation} \begin{center} 
             PRESENTED AT\end{center}\bigskip 
      \begin{center}\begin{large}}{\end{large}\end{center} \end{quotation}}
\def\Acknowledgements{\bigskip  \bigskip \begin{center} \begin{large}
             \bf ACKNOWLEDGEMENTS \end{large}\end{center}}

\begin{document}
\begin{titlepage}

\Title{Measurements of \\
Inclusive b--quark Production at 7 TeV \\
with the CMS Experiment \\}
\vfill
\Author{P. Bellan \\
on behalf of\\
the CMS Collaboration\\}  
\Address{Padova University and INFN -- 35131 Padova, Italy}
\vfill



\begin{Abstract}
\noindent
Measurements performed by the CMS experiment of the cross section for inclusive b-quark production in proton-proton collisions at $\sqrt{s} = 7$ TeV are presented. The measurements are based on different methods, such as inclusive jet measurements with secondary vertex tagging or selecting a sample of events containing jets and at least one muon, where the transverse momentum of the muon with respect to the closest jet axis discriminates b events from the background. The results are compared with predictions based on perturbative QCD calculations at leading and next-to-leading order.
\\
\end{Abstract}
\vfill

\begin{Presented}
Hadron 2011\\
Munich, Deutschland \\
June 13--17, 2011 \\
\end{Presented}
\vfill

\end{titlepage}


\section{Introduction}\label{ref:intro}

\noindent
The study of heavy-quark production in high-energy hadronic interactions
plays a key role in testing next-to-leading order (NLO) 
Quantum Chromodynamics (QCD) calculations. 
In the past, discrepancies were observed between experimental data and theoretical predictions, e.g. at Tevatron~\cite{d0:inclmub, d0:bb, cdf:bsemilept, cdf:b2mud0x} 
and HERA~\cite{h1:openb, h1:muj, zeus:mudij, zeus:bc2mu}. 
Substantial progress has been achieved in the understanding of heavy-quark 
production at Tevatron energies~\cite{Cacciari04}, but large theoretical 
uncertainties still remain, mainly due to the dependence of the calculations on the renormalization and factorization scales.
The observed large scale dependence of the NLO calculations is considered
to be a symptom of large contributions from higher orders: small-$x$
effects~\cite{Collins:1991ty, Catani:1990eg}, where 
$x \sim m_\mathrm{b}/\sqrt{s}$, are possibly 
relevant in the low transverse
momentum (\pt) domain, while multiple-gluon 
radiation leads to large logarithms of \pt$/m_\mathrm{b}$ and may be
important at high \pt\cite{Cacciari:1993mq}. The resummed logarithms of
\pt$/m_\mathrm{b}$ at next-to-leading-logarithmic accuracy have been matched to
the fixed-order NLO calculation for massive
quarks~\cite{Cacciari:1998it}.  At the non-perturbative level, the b--hadron
\pt~spectrum depends strongly on the parametrization of the
fragmentation function~\cite{Frixione:1997ma}.  The b-quark
production cross section has also been studied in the general-mass
variable-flavor-number scheme~\cite{Kniehl:2008zza} and the $k_\mathrm{T}$
factorization QCD approach~\cite{Ryskin:2000bz,Jung:2001rp}.  

Measurements of b--hadron production 
at higher energies than before, provided
by the Large Hadron Collider (LHC), represent an important 
test of the new theoretical calculations just mentioned. 
Measurements of inclusive b--quark production cross section 
require identification of inclusive events in which a 
b-quark has been produced in $pp$ collisions.
In results reported here, the discrimination of the heavy quark events 
has been achieved either exclusively, reconstructing the whole decay 
channel of a B meson, or
inclusively, considering hard jets. In this context, two different tagging techniques
have been applied: reconstruction of a displaced secondary vertex, and 
analysis of the transverse momentum spectrum of
an energetic muon with respect to the closest jet.

\section{The CMS detector}
A detailed description of the CMS detector can be found elsewhere~\cite{JINST}.
Some of the most relevant features for heavy flavor physics are summarized here. 

The core CMS apparatus is a superconducting solenoid, of 6~m internal diameter, 
providing a magnetic field of 3.8~T.  
Within the field volume there are the silicon pixel and strip tracker, the crystal electromagnetic calorimeter and the brass/scintillator hadron calorimeter. 
Muons are detected by three types of gas-ionization detectors embedded in the steel return yoke: Drift Tubes (DT), Cathode Strip Chambers (CSC), and Resistive Plate Chambers (RPC).
The muon detectors cover a pseudorapidity window $|\eta|< 2.4$, where $\eta = - \ln [\tan (\theta / 2)]$, where the polar angle $\theta$ is measured from the $z$-axis, which points along the counterclockwise beam direction. 
The silicon tracker is composed of pixel detectors (three barrel
layers and two forward disks on each side of the detector, made of
66~million $100\times150$~$\mu$m$^2$ pixels) followed by microstrip detectors
(ten barrel layers plus three inner disks and nine forward disks on each side of the
detector, with 10~million strips of pitch between 80 and 184~$\mu$m).
Thanks to the strong magnetic field and the high granularity of the silicon tracker, the
transverse momentum, $p_\mathrm{T}$, of the muons matched to reconstructed
tracks is measured with a resolution of about 1\,\%  for the typical muons used in this analysis.
The silicon tracker also provides the primary vertex position, with $\sim$\,20~$\mu$m accuracy.  

The first level (L1) of the CMS trigger system, composed of custom
hardware processors, uses information from the calorimeters and muon
detectors to select the most interesting events.  The High Level
Trigger (HLT) further decreases the rate before data storage.

\section{Open beauty}
b-­‐jets play a key role in searches
for new physics beyond the Standard Model.
It took a while to fully establish a consistency
between the Tevatron data and pQCD predictions for b--jet production cross section. 
Generally speaking, b--jets cross section measurements are highly non-trivial, and 
sizable uncertainties affect both theory and experiment: on one side, 
one has to deal with a typical multi--scale problem, in which 
the center--of--mass collision energy, mass of the b--quark and the factorization and re--normalization scales are entangled in a subtle way; 
on the other hand, excellent performance of the tracking is required, challenging the detector full potential. 

The CMS Collaboration has published results, 
reviewed in the next sections, concerning 
two complementary measurements of b--jet production cross section.
These results have made use of two different b--tagging techniques, were
performed over $85$ nb$^{-1}$ and $60$ nb$^{-1}$ of 2010 data, respectively~\cite{cms:bJets-pTrel,cms:bJets-SV}, and 
 were obtained using track or particle--flow jets.
Typical values achieved by the CMS detector of the jet resolution and 
the energy scale uncertainty, at the time of these results were reported, 
were about $10-15$\%,  and below $ 3$\%, respectively.

\subsection{b--jets with muons}
The measurement of the integrated
and differential cross section of the reaction
$pp \rightarrow b+X \rightarrow \mu+X$
has been performed on jets coming from b--quarks.
The production of a b--quark decaying semi--leptonically is deduced by the identification of a rather energetic muon inside a jet, where the transverse momentum
relative to the jet axis is quite sizable.
For a muon from a b--decay, the transverse momentum
relative to the jet axis is on average larger 
than when the muon comes from light quarks; through this property it is hence 
possible to discriminate events in which b--quarks were produced.

\begin{figure}[!htb]
 \begin{center}
   \includegraphics[angle=90,width=0.55\textwidth]{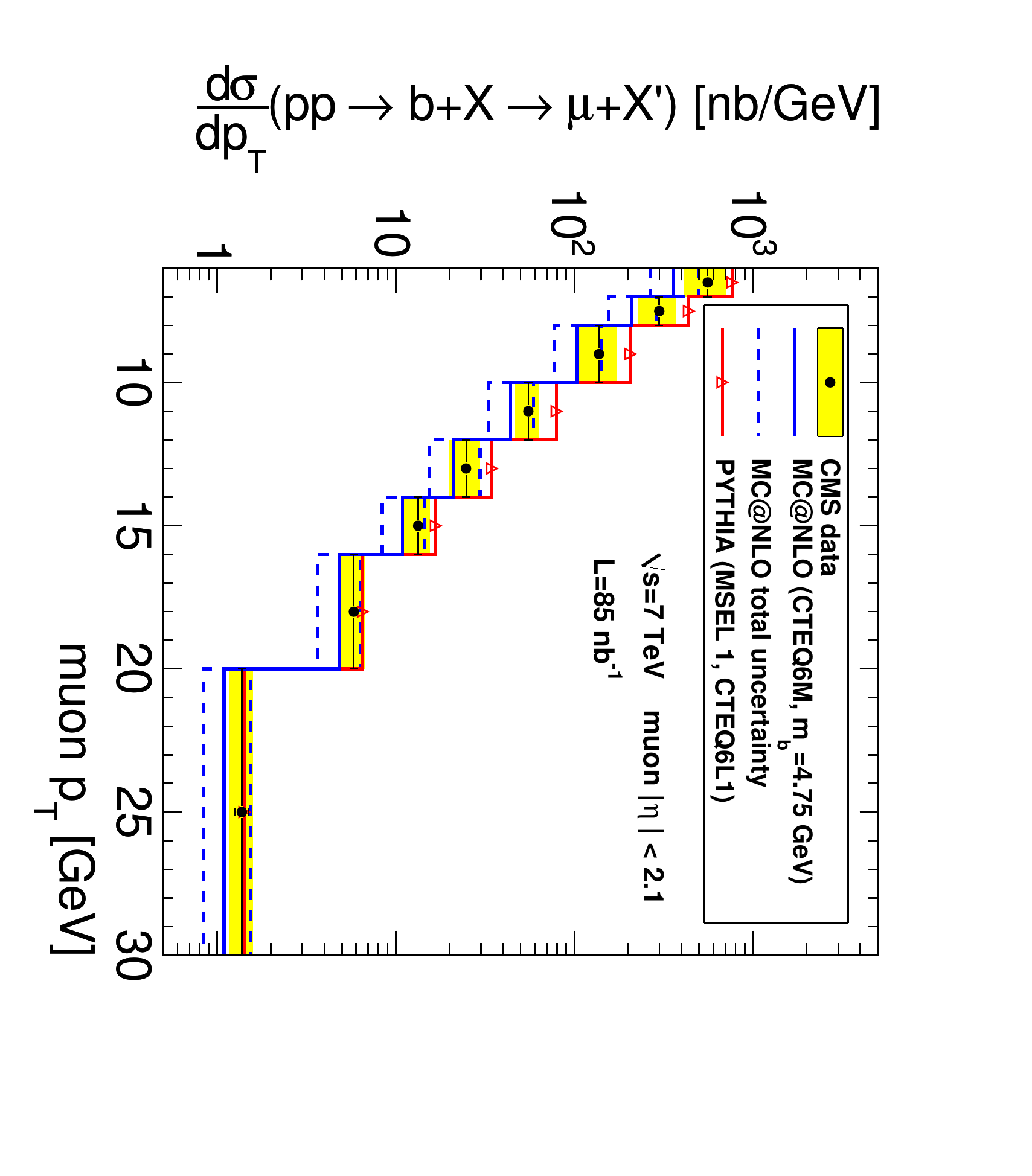}%
   \includegraphics[angle=90,width=0.55\textwidth]{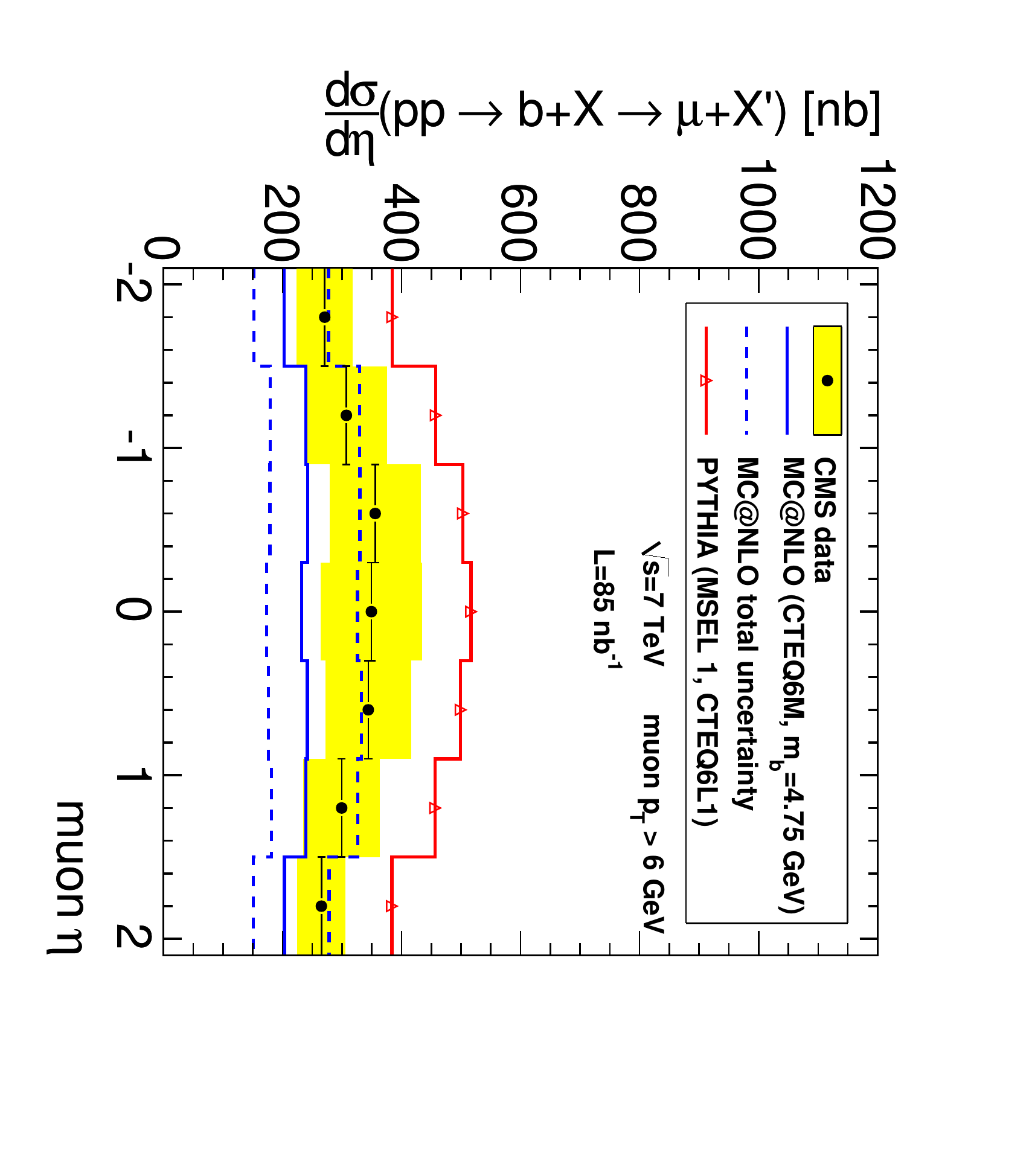}
  \caption{Differential cross section (left)
  $\frac{d\sigma}{d p_{\mathrm{T}}^{\mu}}(pp \rightarrow b+X\rightarrow \mu +X', |\eta^{\mu}|<2.1)$, and   (right) $\frac{d\sigma}{d\eta^{\mu}}(pp\rightarrow \b+X\rightarrow \mu +X', p_{\mathrm{T}}^{\mu}> 6 $ GeV). The two possible muon charges are not distinguished and the process $pp \rightarrow \bar{b}+X \rightarrow \mu +X'$ is included.
    The black points are the CMS measurements. Vertical error bars showing the statistical error are smaller than the point size in most bins, the horizontal bars indicate the bin width. The yellow band shows the quadratic sum of statistical and systematic
    uncertainties. The systematic uncertainty (11\%) of the luminosity
    measurement is not included.
    The solid blue line shows the MC@NLO result and the dashed blue lines illustrate the
    theoretical uncertainty as described in the text.
    The solid red line with dots shows the PYTHIA result.
\label{pTrel-absu}}
 \end{center}
\end{figure}

A binned log-likelihood fit is performed on the spectrum of such a quantity, called ``$p_{\mathrm{T}}^{rel}$'', 
using template distributions provided by the simulation for $b$ and $c$ quarks, and
derived from the data for gluons and lighter quarks.
This latter distribution is dominated by hadrons 
misidentified as muons (mainly decay-in-flight), 
so they are reweighted by the misidentification rate measured in the data.
Considering that the fit is not able to distinguish between light
quark, gluon and charm components, 
these are merged together. 
The b--jet tag efficiency achieved with this technique is about $74$\% at $ p_{\mathrm{T}}^{\mu} \simeq 6$ GeV, and close to $100$\% above $20$ GeV, whereas the contamination is $\sim 7$\% in lowest \pt~bin, asymptotically decreasing
towards $2$\% at high \pt.

The result of the inclusive production cross section for b quarks decaying into muons within
the kinematic range $p_{\mathrm{T}}^{\mu}> 6 $ GeV and $|\eta^{\mu}|<2.1 $ is:
\begin{displaymath}
 \sigma  = 1.32 \pm 0.01\mathrm{(stat)} \pm 0.30 \mathrm{(syst)} \pm 0.15 \mathrm{(lumi)} ~ \mu b
\end{displaymath}
where the first uncertainty is statistical, the second is systematic, and the third is associated with the estimation of the integrated luminosity.
Fig.~\ref{pTrel-absu} shows the measured value of the cross section 
for the reaction $pp \rightarrow b+X \rightarrow \mu+X$, as a function of the muon's \pt~and $y$.
The result includes efficiency of trigger $(88 \pm 5)$\%, 
of muon reconstruction $(94 \pm 3)$\% and muon-jet association $(77 \pm 8)$\%.

The fit stability was tested against variation of the binning, repeating the fits on many simulated pseudo--experiments, cross--checking the results using jets from particle flow and performing the fits on the impact parameter distribution.

\begin{table}[!htb]
 \begin{center}
  \caption{Summary of systematic cross section uncertainties. The systematic uncertainty
    can vary depending on the muon transverse momentum and
    pseudorapidity as indicated by the range.}
  \begin{tabular}{l r}
  \\
   \hline
   source   &  cross section uncertainty (\%)\\
   \hline
   Trigger efficiency                    &     5   \\
   Muon reconstruction efficiency        &     3   \\
   Hadron tracking efficiency            &     2   \\
   b $p_\mathrm{T}^{rel}$  shape uncertainty          &  $\le$ 21   \\
   Background $p_\mathrm{T}^{rel}$  shape uncertainty & 2--14   \\
   Background composition                &  3--6   \\
   Production mechanism                  &  2--5   \\
   Fragmentation                         &  1--4   \\
   Decay                                 &     3   \\
   Underlying event                      &    10   \\
   Luminosity                            &    11   \\
   \hline
  \end{tabular}
  \label{t:systematics}
 \end{center}
\end{table}

Table~\ref{t:systematics} summarizes the systematic uncertainties entailed in this measurement.
The uncertainties dominating the measurement are those coming from the approximate knowledge of the signal and the background $p_\mathrm{T}^{rel}$ shape.

\begin{figure}[htb!]
\centering
\includegraphics[angle=90, width=0.55\textwidth]{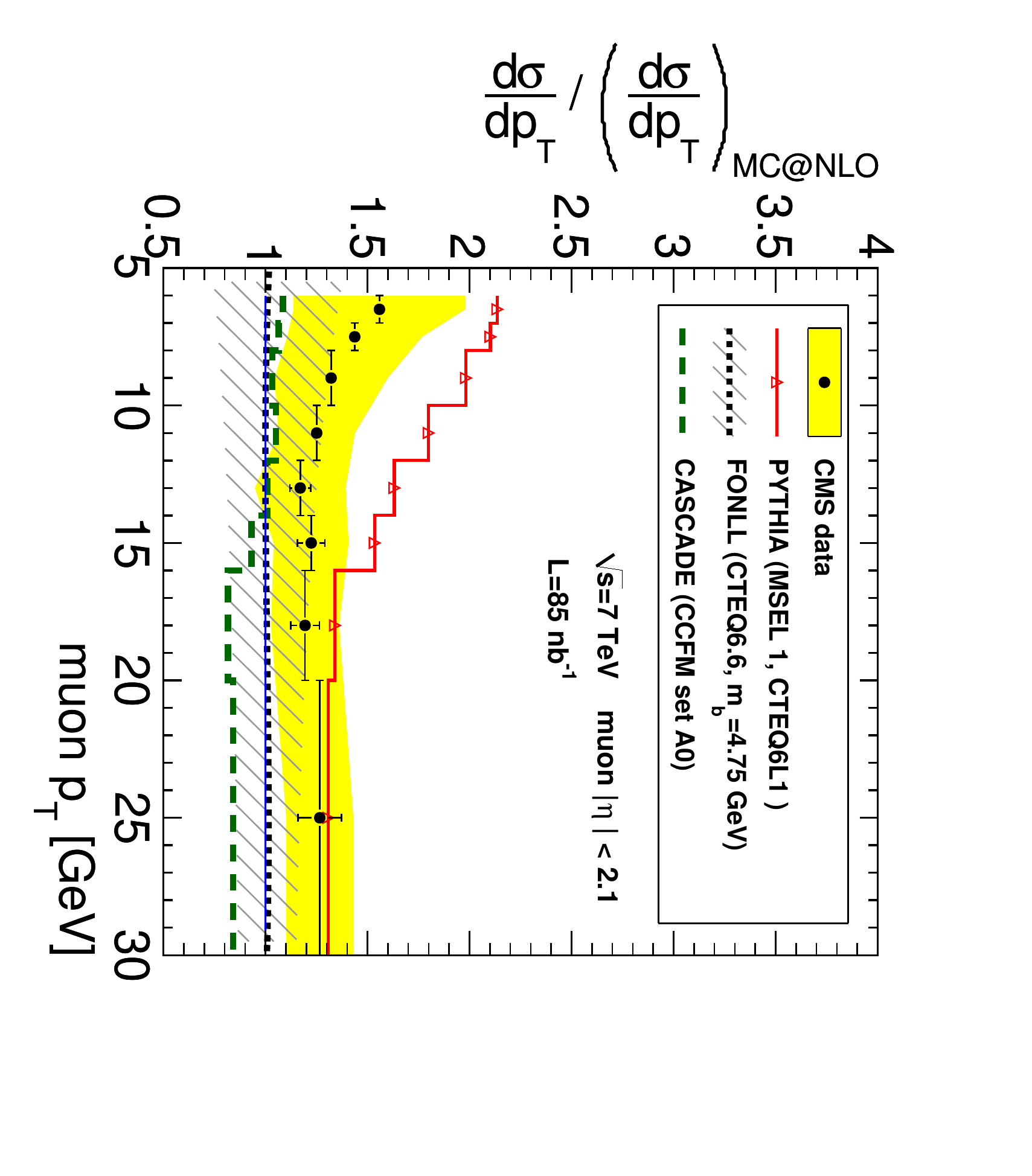}%
\includegraphics[angle=90, width=0.55\textwidth]{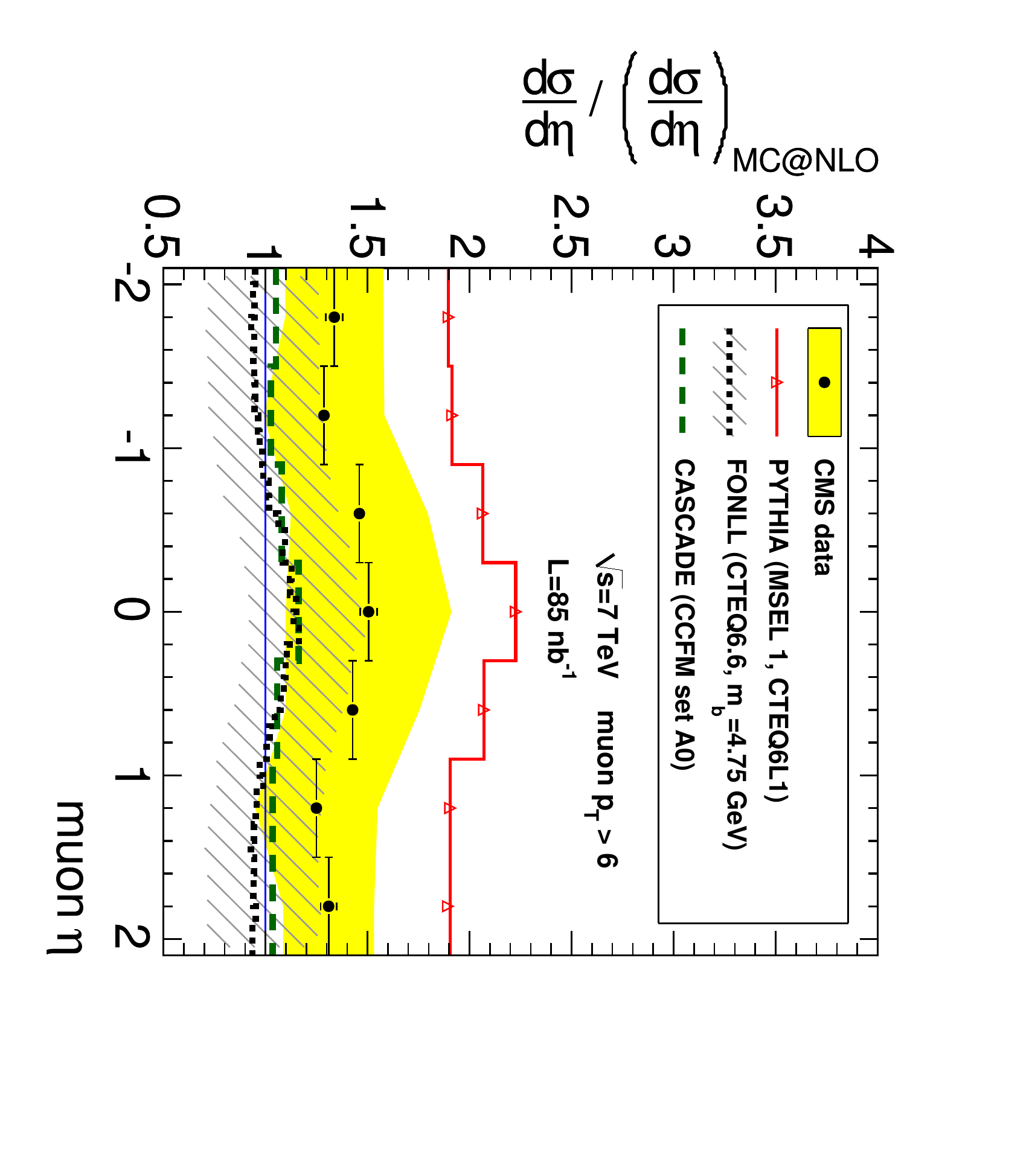}
\caption{
\small{Differential cross section (left)
  $\frac{d\sigma}{d p_{\mathrm{T}}^{\mu}}(pp \rightarrow b+X\rightarrow \mu +X', |\eta^{\mu}|<2.1)$, and   (right) $\frac{d\sigma}{d\eta^{\mu}}(pp\rightarrow b+X\rightarrow \mu +X', p_{\mathrm{T}}^{\mu}> 6 $ GeV), divided by their MC@NLO predictions. The two possible muon charges are not distinguished; the cross section includes the process 
$pp \rightarrow \bar{b} + X \rightarrow \mu  + X$.
The black points are the CMS measurements. Vertical error bars showing the statistical error are smaller than the point size in most bins, the horizontal bars indicate the bin width. The yellow band shows the quadratic sum of statistical and systematic uncertainties. The systematic uncertainty
(11\%) of the luminosity measurement is not included. Superimposed are the FONLL result (black dotted line) with uncertainties (hatched band), the CASCADE result (green, dashed line) and the pythia result (red line with markers), divided by the MC@NLO cross section.}}
\label{fig:pTrel-ratios}
\end{figure}

Fig.~\ref{fig:pTrel-ratios} shows the ratio of the measured differential production cross sections in \pt~and $\eta$ over the predictions of the MC@NLO computation, as well as 
those provided by the Pythia~\cite{bib-PYTHIA}, CASCADE MC~\cite{cascade} and the FONLL~\cite{fonll} calculations.

\subsection{b--tagging with secondary vertices}
\begin{figure}[htb!]
\centering
\includegraphics[width=0.49\textwidth]{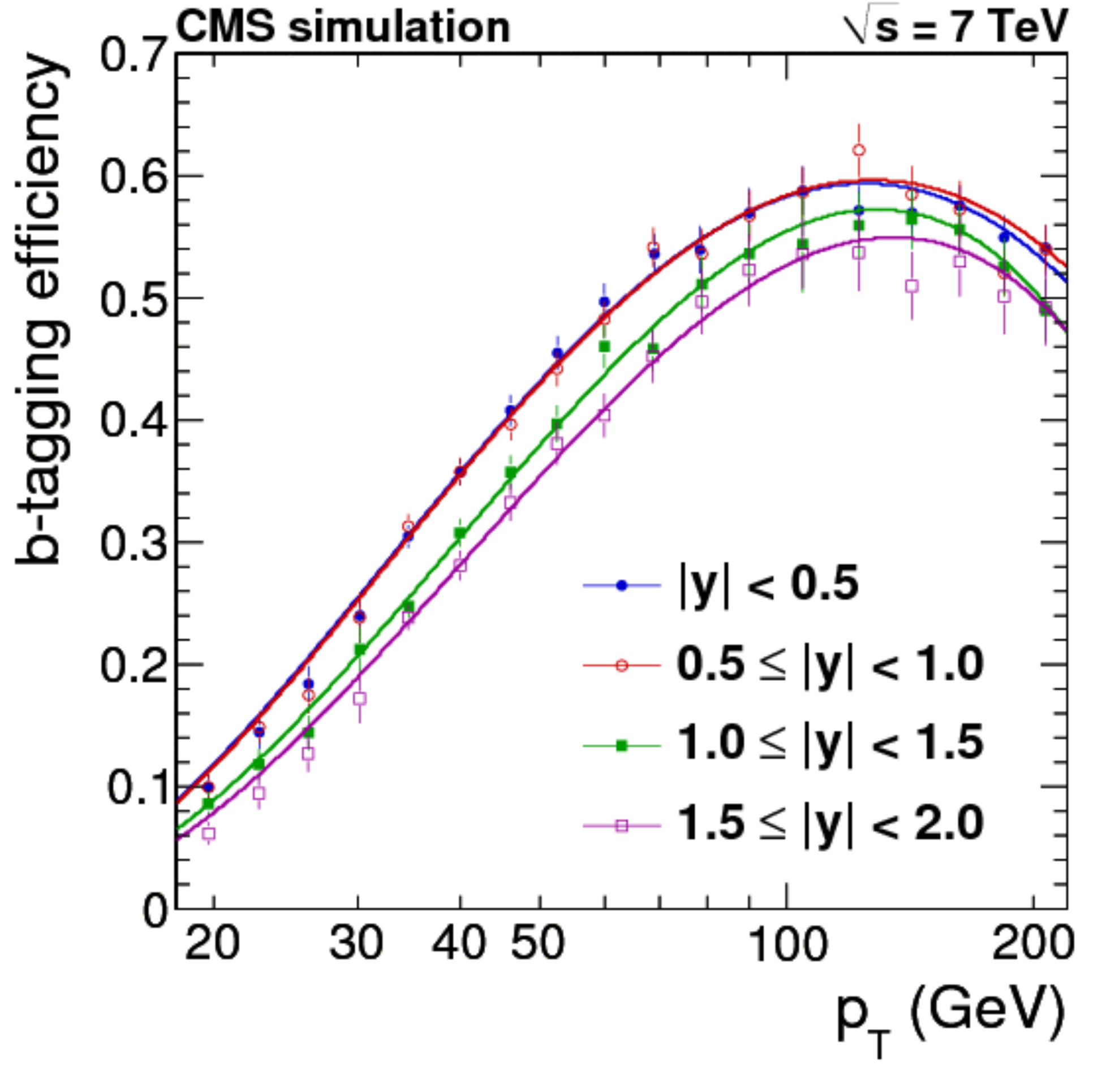}%
\includegraphics[width=0.49\textwidth]{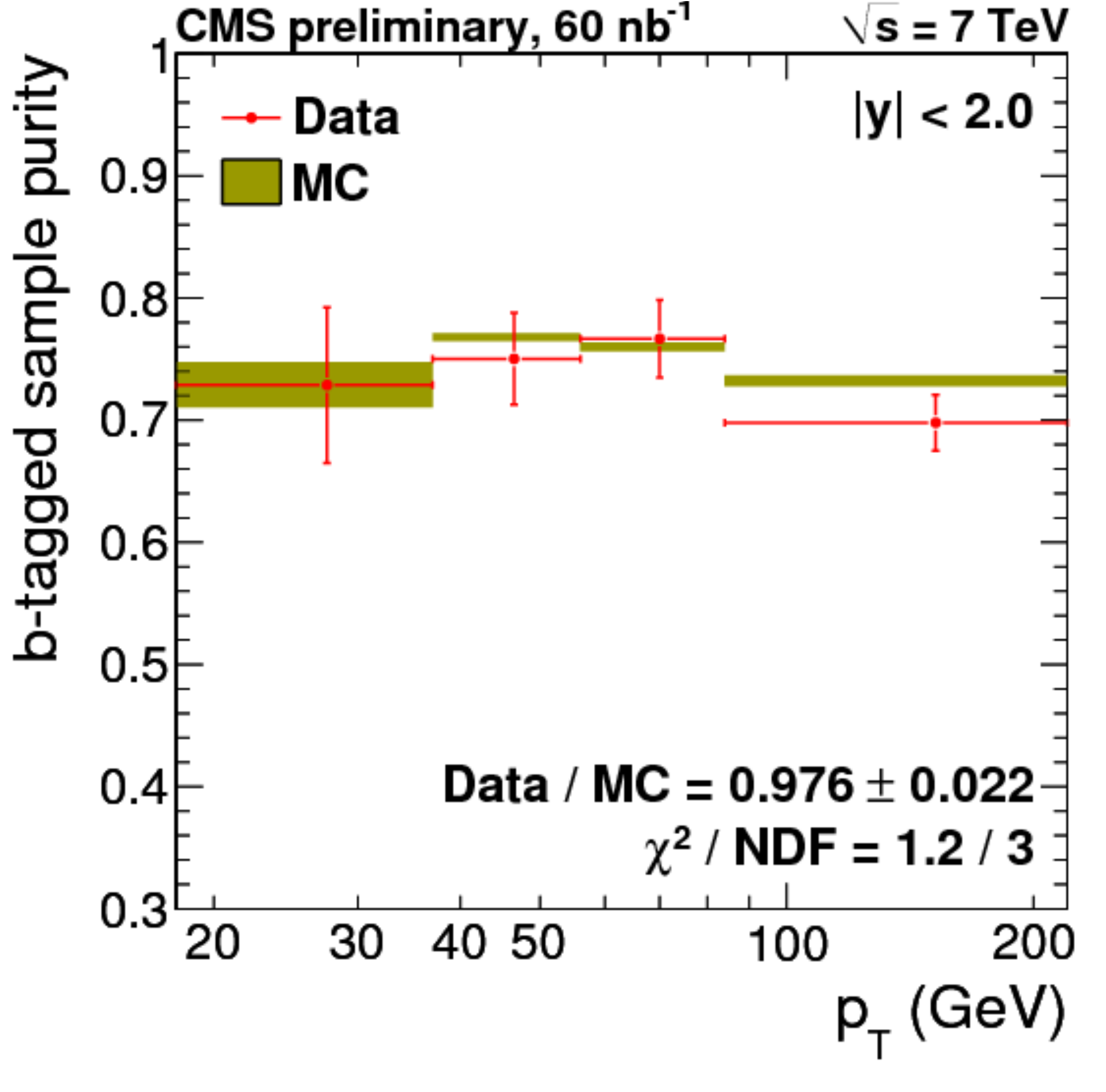}
\caption{
\small{b--tagging efficiency in different rapidity bins, as estimated on simulated events (left). The b--tagged sample purity obtained using fits to secondary vertex mass (right) }}
\label{SV_effipuri}
\end{figure}

The identification of jets coming from the hadronization of b--quarks is possible 
also through the reconstruction of secondary vertices (SV).
Once a displaced SV is reconstructed, different discriminators 
can be used to tag the jet as originating 
from a b--quark. In the analysis presented here, the discriminator adopted is a 
monotonic function of the 3D decay length.
The decay length significance
cut is chosen so that the corresponding
tagging efficiency is about $60$\% at $p_\mathrm{T}^{jet} = 100$ GeV, 
with a  contamination of $\sim0.1$\%.
The b--tagging efficiency and the mistag rates from $c$ or light jets are evaluated from simulated events and constrained by a data/MC scale factor.
The b--tagging efficiency with the selections used in this analysis is between 6\% and 60\% 
at \pt$> 18$ GeV and $|y| < 2.0$. The efficiency rises at higher \pt~as 
the b--hadron proper-time increases.
In order to evaluate the purity of the selected sample, 
a fit to the SV mass distribution is performed, 
taking the shapes from simulated events, and letting free
the relative normalizations for c and b jets, with the (small) contribution from light quarks
fixed to the Monte Carlo expectations (``template fit''). 
The efficiencies estimated from MC, and the estimates of b-tagged sample purity resulting from fits to secondary-vertex mass from data \mbox{are shown in Fig.~\ref{SV_effipuri}.}

The data sample was collected using hadronic triggers with different thresholds of the jet  $p_\mathrm{T}$. To merge them, the individual \pt~spectra of jets have been normalized with the luminosity of their data taking periods, and 
then combined into a single spectrum with the jet \pt~bins corresponding to intervals
where the triggers were fully efficient.
 The overall transverse jet energy range goes from $18$ to 
$300$ GeV, and the measurements have been performed in four $\eta$ intervals. 
The jet energy corrections applied for rapidity dependence, and those for absolute scale and \pt~dependence,  come from real data and simulated events, respectively.

\begin{figure}[h!]
\centering
\includegraphics[width=0.49\textwidth]{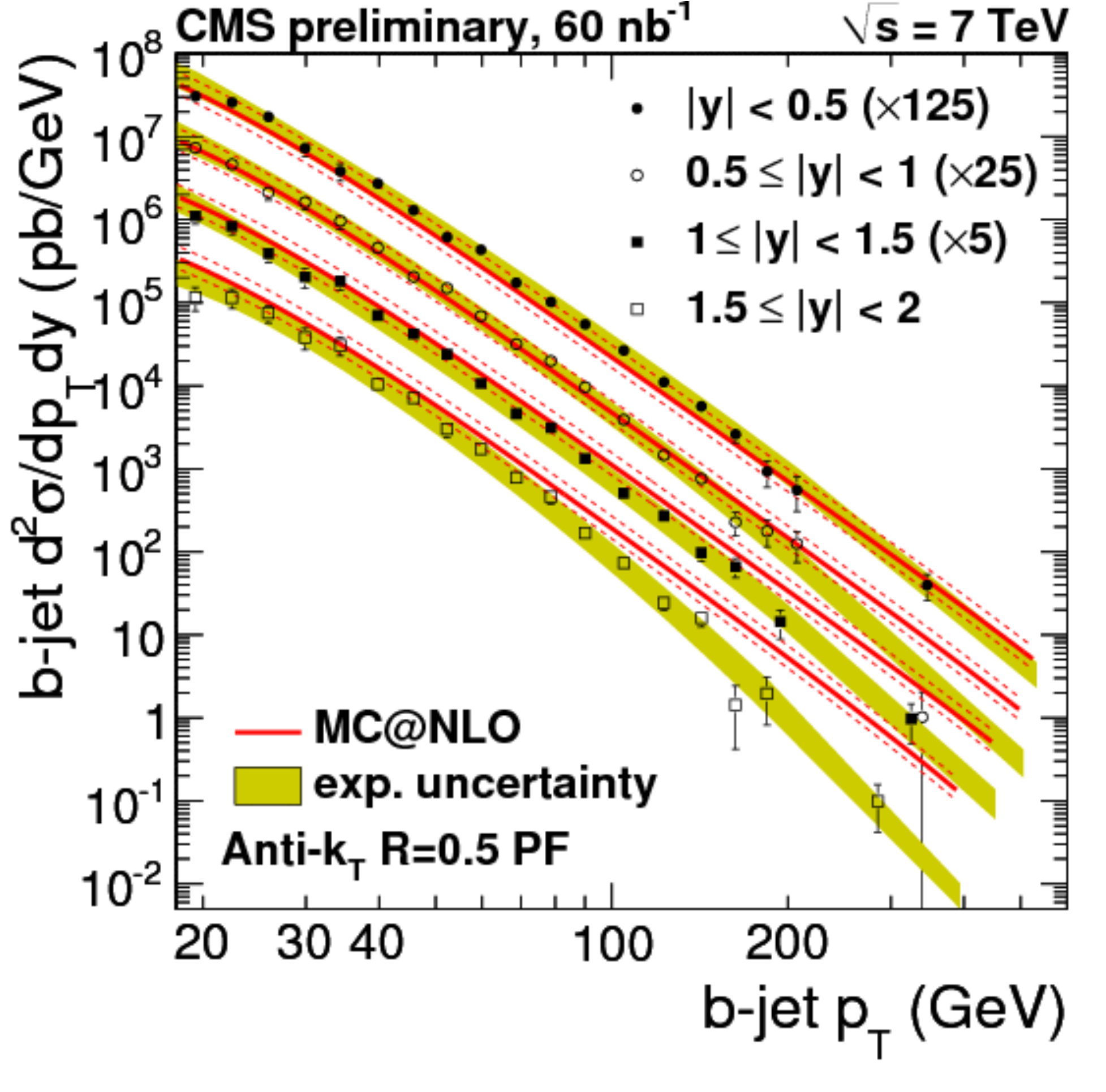}%
\includegraphics[width=0.49\textwidth]{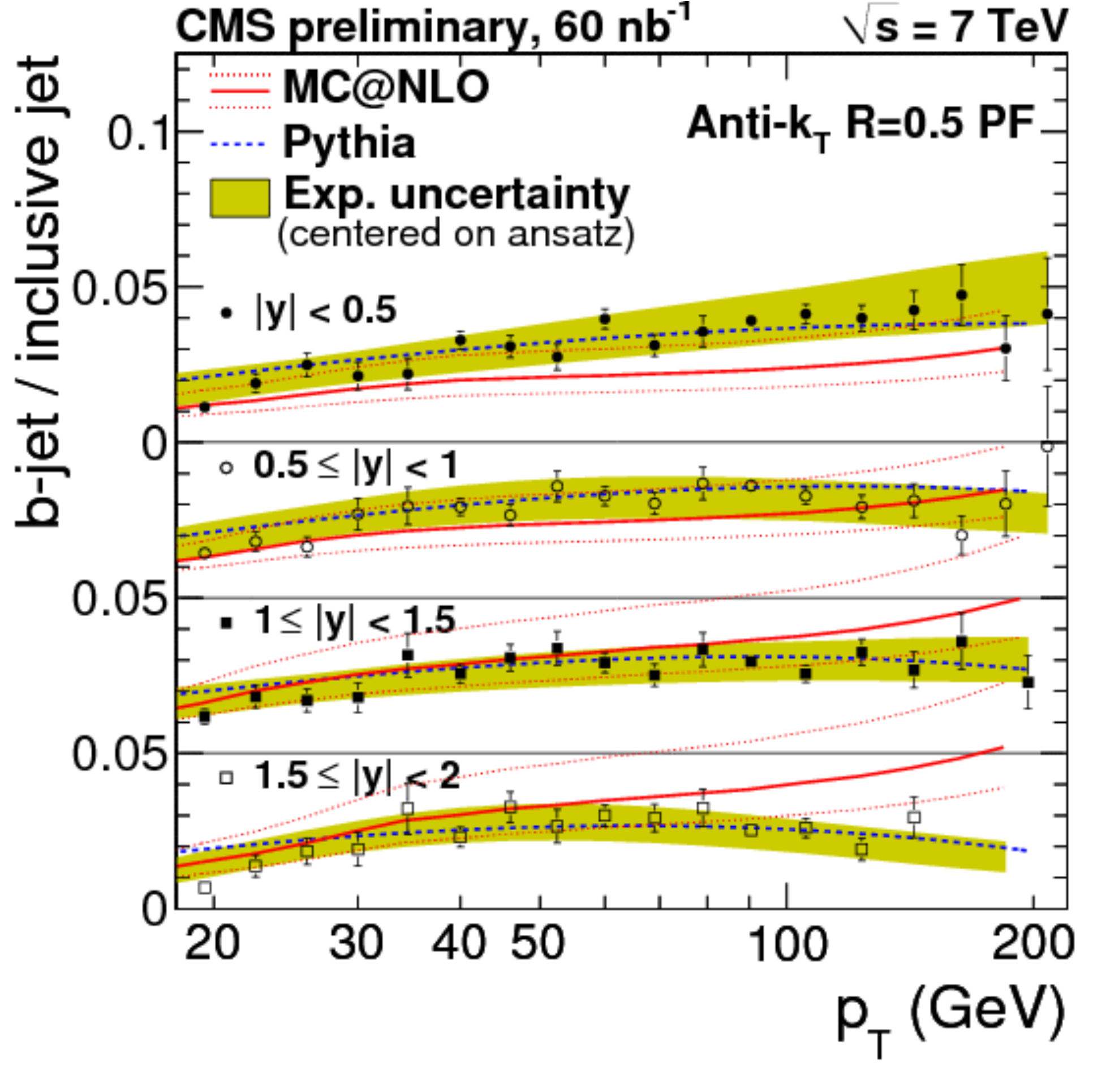}
\caption{
\small{Production cross section measurement for  b--jet as a function of the b--jet transverse momentum 
compared with the MC@NLO predictions (left) and the ratio of the b--jet cross section over the inclusive jets (right).}}
\label{fig:bjetXsec}
\end{figure}

The leading systematic uncertainties affecting the measurements are:
\vspace{-0.2cm}
\begin{itemize}
\item the jet energy scale of b--jets relatively to the inclusive ones ($4–5$\%);
\vspace{-0.2cm}
\item data-driven constraints on b-tagging efficiency ($20$\%);
\vspace{-0.2cm}
\item mistag rate for charm ($3–4$ \%) and for light jets ($1$ –- $10$\%).
\end{itemize}

Fig.~\ref{fig:bjetXsec} shows the results for the production cross section measurements 
for b--jets as a function of the b--jet transverse momentum, compared 
with the MC@NLO predictions, and the ratio with the inclusive jets cross-section.
While the agreement with Pythia and MC@NLO is reasonable, significant differences in shape 
are evident, the simulations predicting more b--jets at high \pt~than what is observed.

\section{Conclusions}
Recent results published by the CMS Collaboration on open beauty production have been summarized, all of them obtained analyzing the $pp$ collision data at $7$ TeV 
collected in the year 2010. 
The studies made with two different techniques for b--tagging have been shown,
together with the comparisons of the production cross section measured in the data with the predictions from the available theoretical models.

\Acknowledgements
The involvement in this work was made possible by the founding
of the INFN, sezione di Padova.
I am grateful to Prof. U. Gasparini for all the sustaining and encouraging, and to 
the entire CMS Padova group for the fruitful collaboration.



\end{document}